%% file: paper.tex
\documentclass[10pt]{article}

\usepackage[T1]{fontenc}
\usepackage[utf8]{inputenc}
\usepackage[english]{babel}
\usepackage{amsmath,latexsym,amsfonts,amsthm,mathtools,dsfont,bm}
\usepackage{graphicx}
\usepackage[top=1.5in, bottom=1.5in, left=1.25in, right=1.25in]{geometry}
\usepackage{booktabs,url}

\usepackage{rotating}
\usepackage{tikz}
\usetikzlibrary{positioning,arrows.meta}

\usepackage{multirow}

\input{mycommands}

\title{{fIRTree: An Item Response Theory modeling \\of fuzzy rating data}}
\author{Antonio Calcagn\`{i} \\\\
		\footnotesize{\sl $^{1}$ University of Padova}\\
		\footnotesize{$\ast$ E-mail: antonio.calcagni@unipd.it}
	}

\date{}

\begin{document}

\maketitle

\begin{abstract}
In this contribution we describe a novel procedure to represent fuzziness in rating scales in terms of fuzzy numbers. Following the rationale of fuzzy conversion scale, we adopted a two-step procedure based on a psychometric model (i.e., Item Response Theory-based tree) to represent the process of answer survey questions. This provides a coherent context where fuzzy numbers, and the related fuzziness, can be interpreted in terms of decision uncertainty that usually affects the rater's response process. We reported results from a simulation study and an empirical application to highlight the characteristics and properties of the proposed approach. \\

\noindent {Keywords:} Fuzzy rating scale; Fuzzy numbers; Item response theory; Decision uncertainty
\end{abstract}

\vspace{2cm}

\section{Introduction}

Over the recent years, fuzzy set theory and fuzzy statistics have been gradually adopted for modeling fuzziness in human rating data \cite{couso2019fuzzy}. Typically, rating data emerge from the measurement of attitudes, motivations, satisfaction, and personality and sociodemographic constructs. In all these cases, as the measurement process involves human raters, data are often affected by fuzziness or imprecision \cite{berglund2012measurement}. Fuzziness in rating data has multiple sources, including the decision uncertainty that affect raters and their response process. Indeed, it is well-known that rating questions like ``I am satisfied with my current work'', using a graded discrete scale from ``strongly disagree'' to ``strongly agree'', entails some degrees of decision uncertainty in human raters. Fuzzy rating scales have been widely adopted with the purpose of quantifying fuzziness from rating data. Mainly, they can be constructed based on two different rationales, one requiring dedicated rating procedures by means of which fuzziness is measured explicitly (e.g., see \cite{de2014fuzzy}) or implicitly (e.g., see \cite{calcagni2014dynamic}), and the other using fuzzy conversion systems through which crisp rating data are converted into fuzzy data either by means of predefined fuzzy systems (e.g., see \cite{vonglao2017application}) or on the basis of statistically-oriented models (e.g., see \cite{yu2009fuzzy,li2016indirect}). Despite their differences, both the approaches aim at modeling fuzziness of rating data.

In this contribution, we describe {fIRTree}, a novel statistically-oriented procedure to model fuzziness from crisp rating data. The aim is to provide an approach to fuzzy rating data that incorporates a stage-wise cognitive formalization of the process that human raters use to answer survey questions. To this end, IRTrees are adopted, a novel class of Item Response Theory (IRT) models that formalize the steps needed by raters to answer multiple choice questions \cite{Boeck_2012,B_ckenholt_2017}. In particular, fIRTree adopts a two-stage modeling system where IRTrees are fit on crisp data first, and then the estimated parameters are used as a bulding block for deriving triangular fuzzy numbers $\text{Trg}(c,l,r,\omega)$ for each rater and item combination. Although several alternatives are available, we resorted to adopt four-parameters triangular fuzzy number $\text{Trg}(c,l,r,\omega)$ because of their flexibility in representing decision rating uncertainty \cite{dombi2018approximations,toth2019applying}. 

The remainder of this contribution is organized as follows. Section 2 describes the rationale of IRTrees and provides the formal notation used throughout this paper. Section 3 presents the proposed fIRTree approach. Section 4 illustrates the results of a simulation study designed to assess the features of fIRTree whereas Section 5 reports a brief case study where fIRTree is applied on empirical data. Finally, Section 6 concludes this paper and provides a summary of the current findings. 

Note that all the materials such as algorithms and datasets used throughout this paper are available to download at \url{https://github.com/antcalcagni/firtree/}.

\section{IRTree models}

\begin{figure}[!h]
	\centering
	\resizebox{7cm}{!}{\input{fig1.tex}}
	\caption{Example of IRTree used to model a common rating scale with five response categories. }
	\label{fig1}
\end{figure}
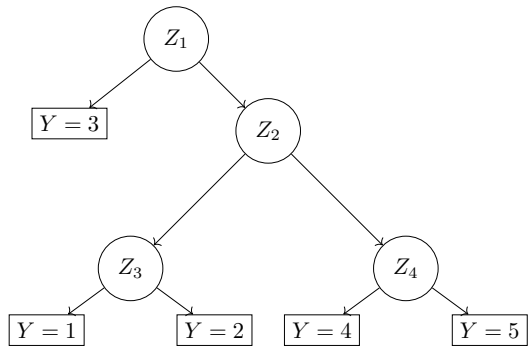

Item Response Theory (IRT) trees are conditional linear models where rating data are modeled in terms of binary trees. They formalize the rating process as a sequence of conditional stages going through the tree to end nodes: in this schema, intermediate nodes represent specific cognitive components of the response process whereas terminal nodes represent the possible outcomes of the decision process. Figure \ref{fig1} depicts the simplest IRT tree model for a rating scales with five choices (e.g., from 1: ``strongly disagree''; to 5: ``strongly agree''). The adoption of this schema would allow for modeling those situations where raters first decide whether or not provide their responses ($Z_1$) and, then, decide on the direction ($Z_2$) and strength of their answers ($Z_3$, $Z_4$). The probability of uncertain responses (i.e., $Y=3$: ``neither agree, neither disagree'') is simply given by the probability to activate the first stage of the decision process, namely $\Prob{Y=0} = \Prob{Z_1;\boldsymbol{\theta}_1}$. On the contrary, the probability of a negative extreme response (i.e., $Y=1$: ``strongly disagree'') is computed as $\Prob{Y=1} =\Prob{Z_1;\boldsymbol{\theta}_1}(1-\Prob{Z_2;\boldsymbol{\theta}_2})(1-\Prob{Z_3;\boldsymbol{\theta}_3})$. In this case, the latent random variables $\{Z_1,\ldots,Z_4\}$ govern the sub-processes of the rater's response. 

As for more general IRT models, the probability to agree $\Prob{Y\in\{4,5\}}$ or disagree $\Prob{Y\in\{1,2\}}$ with an item can be represented as a function of a rater's latent trait and the specific content of the item \cite{Boeck_2012}. More formally, let $i \in \{1,\ldots,I\}$ and $j \in \{1,\ldots,J\}$ be the indices for raters and items, respectively. Then, the final response variable $Y_{ij} \in \{1,\ldots,m,\ldots,M\} \subset\mathbb N$, with $M$ being the maximum number of response categories, can be decomposed into $N$ Boolean variables $Z_{ijn} \in \{0,1\}$, with $n \in \{1,\ldots,N\}$ denoting the nodes of the tree. For a generic rater-item pair $(i,j)$, the IRTree consists of the following equations:
\begin{align}
& \boldsymbol\eta_{i} \sim \mathcall N(\mathbf 0,\boldsymbol{\Sigma}_{\eta})\label{eq1a}\\
& \pi_{ijn} = \Prob{Z_{ijn} = 1; \boldsymbol{\theta}_n} = \frac{\exp\left(\eta_{in}+\alpha_{jn}\right)}{1+\exp\left(\eta_{in}+\alpha_{jn}\right)}\label{eq1c}\\
& Z_{ijn} \sim \mathcall Ber(\pi_{ijn})\label{eq1d}
\end{align}
where $\boldsymbol{\theta}_n = \{\boldsymbol{\alpha}_j, \boldsymbol{\beta}_i\}$, with the arrays $\boldsymbol{\alpha}_j \in \mathbb R^N$ and $\boldsymbol{\eta}_i \in \mathbb R^N$ denoting the easiness of the item and the rater's latent trait. As for any IRT model, latent traits are modeled using a Gaussian distribution centered on zero and with covariance matrix $\boldsymbol{\Sigma}_\eta$. Overall, the probability for a given response category $\Prob{Y=m;\boldsymbol{\theta}}$ is given by recursion from the single-branch probability equation as follows:
\begin{align}
\Prob{Y_{ij}=m} & = \prod_{n=1}^N \Prob{Z_{ijn} = t_{mn};\boldsymbol{\theta}_n}^{t_{mn}} \nonumber\\
& = \prod_{n=1}^N \left(\frac{\exp\left(\eta_{in}+\alpha_{jn}\right)t_{mk}}{1+\exp\left(\eta_{in}+\alpha_{jn}\right)}\right)^{\delta_{mn}}\label{eq2}
\end{align}
where $t_{mn}\in\{0,1,\text{NA}\}$ is the entry of the mapping matrix $T_{M\times N}$ with $t_{mn}=1$ indicating a connection from the $m$-th response category to the $n$-th node, $t_{mn}=0$ or $t_{mn}=\text{NA}$ indicating no connection at all, whereas $\delta_{mn}=0$ if $t_{mn}=\text{NA}$ and $\delta_{mn}=1$ otherwise. IRTree models can be estimated either by means of general methods used for generalized linear mixed models, such as restricted or marginal maximum likelihood \cite{Boeck_2012,de2011estimation}, or using procedures for multidimensional IRT models, such as expectation-maximization algorithms \cite{Jeon_2015}. By and large, IRTrees are flexible enough to model (i) simple situations like those requiring unidimensional latent variables (a single $\eta$ for each node of the tree) or common item effects (a single $\alpha$ for each node of the tree), (ii) more complex scenario involving multidimensional high-order latent variables \cite{Boeck_2012}, and (iii) more general schemata requiring graded rating scales with more or less then five response categories \cite{B_ckenholt_2017}.

\section{fIRTree model for rating data}\label{fIRTmap}

Once an IRTree has been fit to a matrix $\mathbf Y_{I\times J}$ of crisp rating data and IRT parameters $\boldsymbol{\hat \theta}=\{\boldsymbol{\hat \eta},\boldsymbol{\hat \alpha}\}$ have been estimated, fuzzy rating data can be obtained by computing the probabilistic model of the rater's multiverse $\mathcall U_{ij}$. This reflects the heterogeneity in the rater's pattern of responses and models all the possible responses the rater would reach if he or she could repeatedly answer a question in the same time and under the same conditions. In this context, decision uncertainty is reflected by the estimated transition probabilities $\Prob{Z_n}$ ($n=1,\ldots,N$) under the general rule that more certain responses require easier transitions among the nodes of the three. Unlike other fuzzy conversion scales grounded on unidimensional IRT models (e.g., see \cite{yu2009fuzzy}), our approach uses $\mathcall U_{ij}$ in order to map fuzzy numbers to the latent response process underlying the observed crisp responses. In this contribution, we used the four-parameters triangular fuzzy numbers \cite{dombi2018approximations,toth2019applying}: 
\begin{align}
\fuzzyset{A}(y; c,l,r,\omega) & = \left({1+\left(\frac{c-y}{y-l}\right)^\omega}\right)^{-1} \cdot \indicatorFun{y}{l}{c} \label{eq5}\\
& + \left({1+\left(\frac{r-y}{y-c}\right)^{-\omega}}\right)^{-1} \cdot \indicatorFun{y}{c}{r} \nonumber
\end{align}
where $l<c<r$ and $\omega \in \mathbb R^+_{0}$. Unlike ordinary triangular fuzzy numbers, by adding the parameter $\omega$ to intensify ($\omega < 1$) or reduce ($\omega > 1$) the fuzziness of the fuzzy set they allow for a flexible representation of the fuzziness present in rating responses. 

Given a pair $(i,j)$, fIRTree consists of the following steps:
\begin{enumerate}
	\item Get the estimates $\boldsymbol{\hat \eta}_{N\times 1}$ and $\boldsymbol{\hat \alpha}_{N\times 1}$ and plug-in them into Eq. \eqref{eq2} to get $\ProbEst{Y=m}$ for each $m \in \{1,\ldots,M\}$. This is the probabilistic model for the rater's multiverse $\mathcall U_{ij}$.
	\item Compute expected value and variance for $\mathcall U_{ij}$:
	\begin{align}
	& c_{ij} = \sum_{y \in \{1,\ldots,M\}} = ~y\cdot\ProbEst{Y=y} \label{eq3}\\
	& s_{ij} = \sum_{y \in \{1,\ldots,M\}} = ~(y-c_{ij})^2\cdot \ProbEst{Y=y} \label{eq4}
	\end{align}
	\item Compute left and right spreads of triangular fuzzy numbers using the following link equations \cite{Williams_1992}:
	\begin{align}
	& {l_{ij}}=c_{ij}-h_2, \quad {r_{ij}}=c_{ij}-h_2+h_1 \nonumber\\
	& h_1 = \sqrt{3.5v_{ij}-3(c_{ij}-\mu_{ij})^2} \label{eq6}\\
	& h_2 = \frac{1}{2}(h_1+3c_{ij}-3\mu_{ij})\nonumber\\
	& \mu_{ij} = (1+c_{ij}(1/s_{ij}))\big/(2+(1/s_{ij}))\nonumber
	\end{align}
	\item Compute the intensification parameter as:
	\begin{equation}\label{eq7}
	\omega_{ij} = \sum_{m=1}^{M} \ProbEst{Y_{ij}=m}^2
	\end{equation}	
	\item The 4-tuple $\{c_{ij},l_{ij},r_{ij},\omega_{ij}\}$ is used to compute the fuzzy number $\tilde y_{ij}$ associated to the observed rater's response $y_{ij}$.
\end{enumerate} 

Note that Eq. \eqref{eq7} is based on the difficulty of responding to an item, a measure which indicates the presence of uncertainty in the response process \cite{meng2014item}. In this representation, when $\ProbEst{Y_{ij}}$ approaches the uniform distribution, $\omega_{ij}$ will reach its minimum value and the fuzzy set will be intensified in order to reach the maximum fuzziness. By contrast, when $\ProbEst{Y_{ij}}$ approaches a degenerated distribution where a single category has probability equals to one, $\omega_{ij}$ will reach the threshold of one and the fuzzy set will turn to the ordinary triangular shape. Hence, the cases where $\omega_{ij} < 1$ progressively indicate that the rater would be very hesitant in responding and he or she would be expected to show a relatively large uncertainty to make a final response. It should also be noted that fuzzy numbers are defined over the space of the means $\Omega(Y)= (1,M)$ instead of being defined over the discrete domain of the responses $\mathcall{R}(Y)=\{1,\ldots,M\}$. 

\section{Simulation study}

The aim of this brief simulation study is twofold. First, we will evaluate whether fIRTree can accurately recover fuzzy numbers. Second, we will provide an external validity check on the results provided by fIRTree in recovering decision uncertainty from crisp rating data. In particular, we assessed fIRTree using a controlled scenario based on simulated faking data (SGR) \cite{lombardi2015sgr} where distinct levels decision uncertainty were progressively increased from a baseline condition. Since it is well known that faking behaviors in rating situations may increase or decrease the overall decision uncertainty of the rater's response process, simulated faking can serve as a good candidate for studying uncertainty in rating process.\footnote{Faking is a deliberate behavior through which respondents distort their responses towards ones they consider more favorable in order to give overly positive self-descriptions, to simulate physical or psychological symptoms as a way to obtain rewards, or to have access to advantageous work positions \cite{zickar2004uncovering}.} The whole simulation study has been performed on a remote HPC machine based on 16 Intel Xeon CPU E5-2630Lv3 1.80Ghz, 16x4 Gb Ram whereas computations and analyses have been performed in the \texttt{R} framework for statistical analyses.\\

\textit{Data generation model}. Crisp rating data were generated using the simplest IRTree for five-point rating scales with a common latent trait and item parameters (i.e., $\eta$ and $\alpha$ were the same across nodes) and the SGR model \cite{lombardi2015sgr}. The latter allows for modulating the presence of faking patterns in the observed data. In particular, it defines the probability of a rating response $Y_{ij}$ as being formed by a true unobserved component $Y_{ij}^T$ and a faking component $Y_{ij}^F$. Simply stated, the probability of answering an item is defined in terms of a conditional replacement distribution $\Prob{Y_{ij}^F|Y_{ij}^T;\boldsymbol{\theta}}$, which models the hypothesized faking behavior of the respondent. The model $\Prob{Y_{ij}^F|Y_{ij}^T;\boldsymbol{\theta}}$ is defined using a particular discrete beta distribution, which has been parameterized in order to represent a set of possible faking behaviors (e.g., faking-good, faking-bad) as well as their intensity (e.g., extreme, slight, uniform). For further information we refer the reader to \cite{lombardi2015sgr}. For the purpose of this study, we manipulated one of the parameter of $\Prob{Y_{ij}^F|Y_{ij}^T;\boldsymbol{\theta}}$, namely $\pi\in [0,1]$ which indicates the probability that an observed response will be replaced by a faking response.  \\

\textit{Design}. The study involved three factors: (i) $I \in \{50,100,150\}$, (ii) $J \in \{10,20\}$, (iii) ${\pi} \in \{0, 0.25, 0.50, 0.75\}$. They were varied in a complete factorial design with a total of $3\times 2\times 4  = 24$ scenarios. For each combination, $B=1000$ samples were generated which yielded to $1000\times 24 = 24000$ new data as well as an equivalent number of parameters. For each condition, the number of response categories and decision nodes were held fixed ($M=5$, $N=M-1=4$) whereas the IRTree for five-point rating scales was adopted (see Fig. \ref{fig1}). Note that, the scenarios with $\pi=0$ correspond to the baseline condition with no faking perturbation. This case has been used to assess the accuracy of fIRTree to recover fuzzy numbers (first aim of the study).\\

\textit{Measures}. The accuracy of fIRTree to recover the parameters of the fuzzy numbers was measured using the average agreement index: $$\text{PA} = \frac{1}{B} \sum_{b=1}^{B}\left( 1- \| \boldsymbol{\hat \Theta}^{(b)} - \boldsymbol{\Theta}^{(b)}\|^2 \big/ \| \boldsymbol{\Theta}^{(b)}\|^2 \right)$$ which indicates how much the recovered parameters $\boldsymbol{\hat \Theta}^{(b)}$ resemble the true parameters $\boldsymbol{\Theta}^{(b)}$ \cite{timmerman2002three}. Values of PA closed to one indicate higher accuracy of the procedure in recovering the true parameters of the model. We computed PA indices for the parameters $c$, $\omega$, and $r-l$ (total spread). The sensitivity of fIRTree to the distinct levels of decision uncertainty (as created by the SGR model) was instead measured by means of the average Kauffmann index: 
\begin{align*}
K = & \frac{1}{B} \sum_{b=1}^B \Bigg(\frac{2}{\text{card}\left(\fuzzyset{A}(y; c^{(b)},l^{(b)},r^{(b)},\omega^{(b)})\right)} \cdot \sum_{y} |\fuzzyset{A}(y; c^{(b)},l^{(b)},r^{(b)},\omega^{(b)}) - \delta(y)| \Bigg)
\end{align*}
with $\delta(y)=1$ if $\fuzzyset{A}(y; c^{(b)},l^{(b)},r^{(b)},\omega)\geq 0.5$ and zero otherwise. The index is a measure of the fuzziness of a set and it approaches one when the set is maximally fuzzy.\\

\textit{Procedure}. Let $i_h$, $j_t$, $\pi_q$ be distinct levels of factors $I$, $J$, $\pi$. Then, crisp rating data were generated according to the following procedure: 
\begin{itemize}
	\item[(a)] $\boldsymbol\eta_{i_h} \sim \mathcall N(\mathbf 0,\mathbf I)$, $\boldsymbol\alpha_{j_t} \sim \mathcall N(\mathbf 1\alpha_0,\mathbf I\sigma_\alpha)$, with $\alpha_0=-1.75$ and $\sigma_\alpha=0.25$
	\item[(b)] Crisp rating data were simulated as follows: $y_{ij} \sim \mathcall{M}ult(M,\Prob{Y_{i_h,j_t}})$, with $\Prob{Y_{i_h,j_t}}$ being computed via Eq. \eqref{eq2} using $\boldsymbol\eta_{i_h}$ and $\boldsymbol\alpha_{j_t}$, for each $i=1,\ldots,i_h$ and $j=1,\ldots,j_t$. Note that $\Prob{Y_{i_h,j_t}}$ was also used to compute the true triangular fuzzy numbers by means of the procedure described in Section \ref{fIRTmap}.
	\item[(c)] For each $i=1,\ldots,i_h$ and $j=1,\ldots,j_t$, observed crisp data $y_{ij}$ were perturbed by the SGR model using the faking parameter $\pi_q$
	\item[(d)] The generated matrices of unperturbed ($\pi_q=0$) and perturbed ($\pi_q>0$) response data $\mathbf Y_{i_h \times j_t}$ were analysed using fIRTree. The nested decision tree with five response categories was adopted (see Figure \ref{fig1}). Since $\boldsymbol{\alpha}$ and $\boldsymbol{\eta}$ were simulated using the simplest model where latent traits and item parameters are invariant across nodes (e.g., see \cite{Boeck_2012}), an IRTree with a common latent trait and common parameters was defined using the \texttt{IRTrees} R library \cite{Boeck_2012}. The \texttt{glmmTMB} R package \cite{glmmTMB} was used to estimate the model parameters via marginal maximum likelihood. Once estimates were obtained, triangular fuzzy numbers were computed using the procedure described in Section \ref{fIRTmap}, which yielded to four new matrices, namely $\mathbf C_{i_j \times j_t}$, $\mathbf L_{i_j \times j_t}$, $\mathbf R_{i_j \times j_t}$, $\mathbf W_{i_j \times j_t}$.\\
\end{itemize}

\begin{table*}[h!]
	\centering
	\begin{tabular}{llccc}
		\hline
		&& $C$ & $R-L$ & $W$ \\ 
		\hline
		\multirow{3}{*}{$J=10$}
		&$I=50$ & 0.779 (0.088) & 0.897 (0.038) & 0.965 (0.013) \\ 
		&$I=150$ & 0.839 (0.027) & 0.916 (0.017) & 0.981 (0.003) \\ 
		&$I=500$ & 0.851 (0.016) & 0.921 (0.011) & 0.983 (0.002) \\ 
		&&&&\\
		\multirow{3}{*}{$J=20$}
		&$I=50$ & 0.805 (0.078) & 0.908 (0.035) & 0.965 (0.014) \\ 
		&$I=150$ & 0.883 (0.028) & 0.942 (0.014) & 0.985 (0.004) \\ 
		&$I=500$ & 0.9 (0.018) & 0.951 (0.008) & 0.988 (0.002) \\ 
		\hline
	\end{tabular}
	\caption{Simulation study: Accuracy of fIRTree to recover fuzzy numbers computed in terms of PA index \cite{timmerman2002three}. Values of PA closed to one indicate that the model accurately recover the true parameters of the fuzzy numbers. Standard deviation over $B=1000$ samples is reported in parenthesis for each scenario. Note that $C$ indicates modes of the fuzzy sets, $R-L$ indicates the total spread of the set, whereas $W$ stands for the intensification parameter.} 
	\label{tab:sim_1}
\end{table*}

\begin{table*}[h!]
	\centering
	\begin{tabular}{llcccc}
		\hline
		&& $\pi=0$ & $\pi=0.25$ & $\pi=0.5$ & $\pi=0.75$ \\ 
		\hline
		\multirow{3}{*}{$J=10$}
		&$I=50$ & 0.617 (0.075) & 0.724 (0.041) & 0.784 (0.019) & 0.815 (0.01) \\ 
		&$I=150$ & 0.602 (0.058) & 0.716 (0.026) & 0.779 (0.01) & 0.812 (0.006) \\ 
		&$I=500$ & 0.603 (0.062) & 0.717 (0.028) & 0.78 (0.008) & 0.813 (0.004) \\
		&&&&&\\
		\multirow{3}{*}{$J=20$}
		&$I=50$ & 0.613 (0.079) & 0.72 (0.045) & 0.781 (0.022) & 0.814 (0.01) \\ 
		&$I=150$ & 0.599 (0.062) & 0.713 (0.03) & 0.776 (0.011) & 0.811 (0.006) \\ 
		&$I=500$ & 0.6 (0.066) & 0.714 (0.032) & 0.777 (0.011) & 0.812 (0.004) \\ 
		\hline
	\end{tabular}
	\caption{Simulation study: Kauffman index of the computed fuzzy numbers as a function of the intensity of the faking perturbation $\pi$. Values of the index closed to one indicate a higher degree of fuzziness in the fuzzy sets. Standard deviation over $B=1000$ samples is reported in parenthesis for each scenario.} 
	\label{tab:sim_2}
\end{table*}

\textit{Results}. Tables \ref{tab:sim_1}-\ref{tab:sim_2} show the results of the simulation study. With regards to the first aim of the study, the accuracy of fIRTree in recovering the true parameters of fuzzy numbers was higher in all the conditions. As expected, the accuracy increased as a function of sample size $I$ and number of items $J$. Overall, fIRTree showed good accuracies in recovering the true parameters of the fuzzy numbers. With regards to the second aim of the study, when compared to the baseline with no faking condition ($\pi=0$ in Table \ref{tab:sim_2}), fuzzy numbers computed via fIRTree behave according to the faking perturbation. As increasing in faking perturbation indicate higher decision uncertainty in rating data, fuzzy numbers progressively increased their fuzziness in the expected direction. Importantly, these results are unaffected by sample size $i_h$ or number of items $j_t$. All in all, these results suggest that fIRTree accurately recovered fuzzy numbers and decision uncertainty when this is present in rating data.

\section{Case study}

\begin{figure}[h!]
	\centering
	\resizebox{7cm}{!}{\input{fig2.tex}}
	\caption{Case study: IRTree used for the six-point rating scale. }
	\label{fig2}
\end{figure}
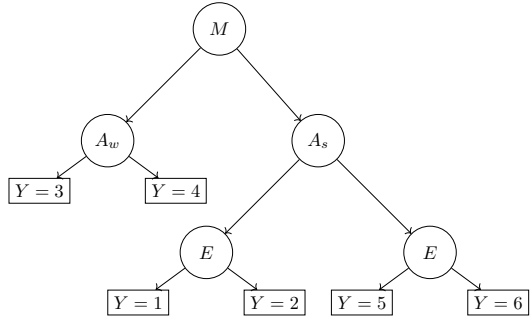

In this section we will shortly describe the application of fIRTree on a real case study involving rating data from the Personality Need for Structure (PNS) questionnaire \cite{neuberg1993personal}. The questionnaire includes 11 items rated on a six-point rating scale (from ``strongly disagree'' to ``{strongly agree}'') and it was designed to measure a person's desire for certainty and decisiveness. The sample consists of $I=497$ respondents, 58\% women, 32.4 years old on average ($SD=11.3$, range = $18-70$), all of them living the United States. Data were originally collected by \cite{B_ckenholt_2017}. A subset of $J=5$ items was selected for the sake of illustration.\footnote{The items were as follows: 1. \textit{It upsets me to go into a situation without knowing what I can expect from it}, 2. \textit{I enjoy having a clear and structured mode of life}, 3. \textit{I like to have a place for everything and everything in its place}, 4. \textit{I don’t like situations that are uncertain}, 5. \textit{I find that a consistent routine enables me to enjoy life more}.} The first step of the analysis consisted in the definition of an IRTree model for the six-point rating scale. To this end, we used the tree suggested by \cite{Berkachy_2019} (see Figure \ref{fig2}). It formally represents a response model where (i) the node $M$ determines whether the respondent has a weak $A_w$ or strong $A_s$ attitude/trait toward the item, (ii) the node $A_s$ activates the strength/extremity $E$ of the rating response, by selecting either low values or higher values of the scale.

\begin{table*}[h!]
	\centering
	\begin{tabular}{cccccccccc}
		\hline
		&& \multicolumn{2}{c}{$M$} & \multicolumn{2}{c}{$A_w$} & \multicolumn{2}{c}{$A_s$} & \multicolumn{2}{c}{$E$}\\ \cmidrule(lr){3-4} \cmidrule(lr){5-6} \cmidrule(lr){7-8} \cmidrule(lr){9-10}
		&& $\hat{\theta}$ & $\sigma_{\hat{\theta}}$ & $\hat{\theta}$ & $\sigma_{\hat{\theta}}$ & $\hat{\theta}$ & $\sigma_{\hat{\theta}}$ & $\hat{\theta}$ & $\sigma_{\hat{\theta}}$ \\ 
		\hline
		&$\alpha_{1}$ & -1.19 & 0.12 & -0.40 & 0.13 & -1.04 & 0.25 & 0.33 & 0.20 \\ 
		&$\alpha_{2}$ & -0.88 & 0.11 & 0.53 & 0.13 & 0.79 & 0.22 & 0.05 & 0.18 \\ 
		&$\alpha_{3}$ & -0.56 & 0.10 & 0.25 & 0.14 & -0.18 & 0.20 & 0.47 & 0.17 \\ 
		&$\alpha_{4}$ & -1.50 & 0.12 & 0.46 & 0.12 & 0.51 & 0.26 & 0.06 & 0.22 \\ 
		&$\alpha_{5}$ & -0.71 & 0.11 & 0.05 & 0.13 & -0.28 & 0.20 & 0.04 & 0.17 \\ 
		\hline
	\end{tabular}
	\caption{Case study: Estimates ($\hat\theta$) and standard errors ($\sigma_{\hat{\theta}}$) for item parameters $\boldsymbol{\hat \alpha}_{J\times N}$.} 
	\label{tab:cs1}
\end{table*}

\begin{table*}[h!]
	\centering
	\begin{tabular}{ccccc|c}
		\hline
		& $\eta_{1}$ & $\eta_{2}$ & $\eta_{3}$ & $\eta_{4}$ & $\hat\sigma_\eta$ \\ 
		\hline
		$\eta_{1}$ & 1.00 &  &  &  & 0.75 \\ 
		$\eta_{2}$ & -0.20 & 1.00 &  &  & 0.95 \\ 
		$\eta_{3}$ & 0.15 & 0.94 & 1.00 &  & 1.16 \\ 
		$\eta_{4}$ & -0.76 & 0.41 & 0.12 & 1.00 & 0.29 \\ 
		\hline
	\end{tabular}
	\caption{Case study: Estimated correlation matrix and standard deviations ($\hat{\sigma}_\eta$) for the latent traits $\boldsymbol{\hat \eta}_{I\times N}$.} 
	\label{tab:cs2}
\end{table*}

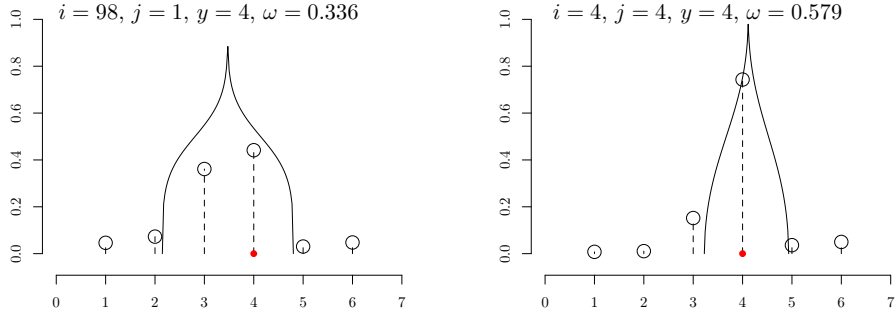
\begin{figure}[h!]
	%\hspace{-0.75cm}
	\centering
	\resizebox{13cm}{!}{\input{fig3.tex}}
	\caption{Case study: Estimated probabilistic model $\{\ProbEst{Y=m}\}_{m=1}^M$ (black dashed vertical lines) for two rater's multiverses on two items and the associated fuzzy number (black dashed curves). The red circle indicates the final crisp responses $y_{ij}$. Note that probability masses and fuzzy membership functions are overlapped over the same domain $\Omega(Y)$ for the sake of comparison.}
	\label{fig3}
\end{figure}

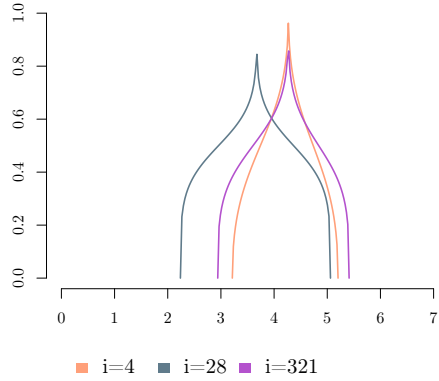
\begin{figure}[h!]
	\centering
	\resizebox{7cm}{!}{\input{fig4.tex}}
	\caption{Case study: Estimated fuzzy numbers for three raters on item $j=2$.}
	\label{fig4}
\end{figure}

The tree with $N=4$ was defined using the \texttt{R} library \texttt{irtree} \cite{Boeck_2012}. Next, the IRTree was fit on the data matrix $\mathbf Y_{497\times 5}$ by means of the \texttt{R} library \texttt{glmmTMB} \cite{glmmTMB}. The estimated parameters $\boldsymbol{\hat \eta}_{I\times N}$ and $\boldsymbol{\hat \alpha}_{J\times N}$ (see Tables \ref{tab:cs1}-\ref{tab:cs2}) were then used to compute the estimated probabilistic models of the multiverses $\{\ProbEst{Y=m}\}_{m=1}^M$ for each rater $i=1,\ldots,I$ and item $j=1,\ldots J$. Finally, fuzzy rating data where obtained using the procedure described in Section \ref{fIRTmap}. Figures \ref{fig3}-\ref{fig4} shows some of the estimated fuzzy numbers. In particular, Figure \ref{fig3} (left panel) shows a prototypical case in which the final crisp response $y_{98,4}=4$ does not correspond to the mode of the estimated fuzzy number. This can happen because fuzzy numbers encapsulate the decision uncertainty that affected the rater's response process, a feature which would otherwise be lost if crisp responses were solely considered. On the contrary, Figure \ref{fig3} (right panel) depicts a situation with lower degree of decision uncertainty.

\section{Conclusion}

In this contribution we presented fIRTree, a novel procedure to represent human rating data in terms of fuzzy numbers. Similarly for other procedures aiming at turning crisp ratings into fuzzy ratings (e.g., see \cite{yu2009fuzzy}), our approach is based on a two-step rationale through which a psychometric rating model (IRtree) is defined and fit on empirical data first, and then the estimated parameters are used to compute 4-tuple triangular fuzzy numbers $\text{Trg}(c,l,r,\omega)$ for each rater and item combination. Although other fuzzy numbers can be adopted to this purpose (e.g., trapezoidal, beta, gaussian), $\text{Trg}(c,l,r,\omega)$ are flexible enough to represent the fuzziness of the rater's response process \cite{dombi2018approximations,toth2019applying}. A simulation study has been designed to assess the accuracy of fIRTree to recover fuzziness of rating data and a case study has been used to illustrate how fIRTree can be applied on empirical cases. 

We believe that fIRTree can improve the way to get fuzzy numbers from rating data. fIRTree does not require dedicated rating tools or measurement settings and it can directly be applied on crisp rating data collected using standard formats of rating scales. Despite its simplicity, it provides a coherent meaning system to compute and interpret fuzzy numbers in measurement contexts involving human ratings.

%\clearpage
\bibliographystyle{plain}
\bibliography{biblio}

\end{document}

%% file: mycommands.tex
\newcommand{\Prob}[1]{\mathbb{P}(#1)}
\newcommand{\ProbEst}[1]{\widehat{\mathbb{P}}(#1)}

\newcommand{\fuzzyset}[1]{\xi_{\widetilde{#1}}}

\newcommand{\indicatorFun}[3]{\mathds{1}_{(#2,#3)}(#1)} %it requires \usepackage{dsfont}
\newcommand{\digamma}[1]{\Psi\left(#1\right)}

\DeclareMathAlphabet{\mathcall}{OMS}{zplm}{m}{n}

%% file: fig1.tex
\begin{tikzpicture}[auto,vertex1/.style={draw,circle},vertex2/.style={draw,rectangle}]
			\node[vertex1,minimum size=1cm] (eta1) {$Z_1$};
			\node[vertex1,minimum size=1cm,below right=1cm of eta1] (eta3) {$Z_{2}$};
			\node[vertex2,below left=1cm of eta1] (Y3) {$Y=3$};
			\node[vertex1,minimum size=1cm,below right=2cm of eta3] (eta3a) {$Z_{4}$};
			\node[vertex1,minimum size=1cm,below left=2cm of eta3] (eta3b) {$Z_{3}$};
			\node[vertex2,below left=0.5cm of eta3b] (Y1) {$Y=1$}; \node[vertex2,below right=0.5cm of eta3b] (Y2) {$Y=2$};
			\node[vertex2,below left=0.5cm of eta3a] (Y5) {$Y=4$}; \node[vertex2,below right=0.5cm of eta3a] (Y6) {$Y=5$};
			
			\draw[->] (eta1) -- node[right=0.5 of eta1] {} (eta3);
			\draw[->] (eta1) -- node[right=0.5 of eta1] {} (Y3); 
			\draw[->] (eta3) -- node[right=0.5 of eta1] {} (eta3a);
			\draw[->] (eta3) -- node[right=0.5 of eta1] {} (eta3b);
			\draw[->] (eta3b) -- node[right=0.5 of eta1] {} (Y1); \draw[->] (eta3b) -- node[right=0.5 of eta1] {} (Y2);
			\draw[->] (eta3a) -- node[right=0.5 of eta1] {} (Y5); \draw[->] (eta3a) -- node[right=0.5 of eta1] {} (Y6);
			\end{tikzpicture}

%% file: fig2.tex
\begin{tikzpicture}[auto,vertex1/.style={draw,circle},vertex2/.style={draw,rectangle}]
			\node[vertex1,minimum size=1cm] (eta1) {$M$};
			\node[vertex1,minimum size=1cm,below left=2cm of eta1] (eta2) {$A_w$};
			\node[vertex1,minimum size=1cm,right=3cm of eta2] (eta3) {$A_s$};
			\node[vertex2,below left=0.5cm of eta2] (Y3) {$Y=3$}; \node[vertex2,below right=0.5cm of eta2] (Y4) {$Y=4$};
			\node[vertex1,minimum size=1cm,below right=2cm of eta3] (eta3a) {$E$};
			\node[vertex1,minimum size=1cm,below left=2cm of eta3] (eta3b) {$E$};
			\node[vertex2,below left=0.5cm of eta3b] (Y1) {$Y=1$}; \node[vertex2,below right=0.5cm of eta3b] (Y2) {$Y=2$};
			\node[vertex2,below left=0.5cm of eta3a] (Y5) {$Y=5$}; \node[vertex2,below right=0.5cm of eta3a] (Y6) {$Y=6$};
			
			\draw[->] (eta1) -- node[right=0.5 of eta1] {} (eta2);
			\draw[->] (eta1) -- node[right=0.5 of eta1] {} (eta3);
			\draw[->] (eta2) -- node[right=0.5 of eta1] {} (Y3); \draw[->] (eta2) -- node[right=0.5 of eta1] {} (Y4);
			\draw[->] (eta3) -- node[right=0.5 of eta1] {} (eta3a);
			\draw[->] (eta3) -- node[right=0.5 of eta1] {} (eta3b);
			\draw[->] (eta3b) -- node[right=0.5 of eta1] {} (Y1); \draw[->] (eta3b) -- node[right=0.5 of eta1] {} (Y2);
			\draw[->] (eta3a) -- node[right=0.5 of eta1] {} (Y5); \draw[->] (eta3a) -- node[right=0.5 of eta1] {} (Y6);
			\end{tikzpicture}

%% file: fig3.tex
% Created by tikzDevice version 0.12.3.1 on 2021-01-10 19:20:15
% !TEX encoding = UTF-8 Unicode
\begin{tikzpicture}[x=1pt,y=1pt]
\definecolor{fillColor}{RGB}{255,255,255}
\path[use as bounding box,fill=fillColor,fill opacity=0.00] (0,0) rectangle (325.21,289.08);
\begin{scope}
\path[clip] ( 49.20, 61.20) rectangle (300.01,239.88);
\definecolor{drawColor}{RGB}{0,0,0}

\path[draw=drawColor,line width= 0.4pt,line join=round,line cap=round] ( 91.67, 82.99) circle (  4.50);

\path[draw=drawColor,line width= 0.4pt,line join=round,line cap=round] (124.84, 87.16) circle (  4.50);

\path[draw=drawColor,line width= 0.4pt,line join=round,line cap=round] (158.02,132.62) circle (  4.50);

\path[draw=drawColor,line width= 0.4pt,line join=round,line cap=round] (191.20,145.26) circle (  4.50);

\path[draw=drawColor,line width= 0.4pt,line join=round,line cap=round] (224.37, 80.49) circle (  4.50);

\path[draw=drawColor,line width= 0.4pt,line join=round,line cap=round] (257.55, 83.22) circle (  4.50);
\end{scope}
\begin{scope}
\path[clip] (  0.00,  0.00) rectangle (325.21,289.08);
\definecolor{drawColor}{RGB}{0,0,0}

\path[draw=drawColor,line width= 0.4pt,line join=round,line cap=round] ( 58.49, 61.20) -- (290.73, 61.20);

\path[draw=drawColor,line width= 0.4pt,line join=round,line cap=round] ( 58.49, 61.20) -- ( 58.49, 55.20);

\path[draw=drawColor,line width= 0.4pt,line join=round,line cap=round] ( 91.67, 61.20) -- ( 91.67, 55.20);

\path[draw=drawColor,line width= 0.4pt,line join=round,line cap=round] (124.84, 61.20) -- (124.84, 55.20);

\path[draw=drawColor,line width= 0.4pt,line join=round,line cap=round] (158.02, 61.20) -- (158.02, 55.20);

\path[draw=drawColor,line width= 0.4pt,line join=round,line cap=round] (191.20, 61.20) -- (191.20, 55.20);

\path[draw=drawColor,line width= 0.4pt,line join=round,line cap=round] (224.37, 61.20) -- (224.37, 55.20);

\path[draw=drawColor,line width= 0.4pt,line join=round,line cap=round] (257.55, 61.20) -- (257.55, 55.20);

\path[draw=drawColor,line width= 0.4pt,line join=round,line cap=round] (290.73, 61.20) -- (290.73, 55.20);

\node[text=drawColor,anchor=base,inner sep=0pt, outer sep=0pt, scale=  1.00] at ( 58.49, 39.60) {0};

\node[text=drawColor,anchor=base,inner sep=0pt, outer sep=0pt, scale=  1.00] at ( 91.67, 39.60) {1};

\node[text=drawColor,anchor=base,inner sep=0pt, outer sep=0pt, scale=  1.00] at (124.84, 39.60) {2};

\node[text=drawColor,anchor=base,inner sep=0pt, outer sep=0pt, scale=  1.00] at (158.02, 39.60) {3};

\node[text=drawColor,anchor=base,inner sep=0pt, outer sep=0pt, scale=  1.00] at (191.20, 39.60) {4};

\node[text=drawColor,anchor=base,inner sep=0pt, outer sep=0pt, scale=  1.00] at (224.37, 39.60) {5};

\node[text=drawColor,anchor=base,inner sep=0pt, outer sep=0pt, scale=  1.00] at (257.55, 39.60) {6};

\node[text=drawColor,anchor=base,inner sep=0pt, outer sep=0pt, scale=  1.00] at (290.73, 39.60) {7};

\path[draw=drawColor,line width= 0.4pt,line join=round,line cap=round] ( 49.20, 75.70) -- ( 49.20,233.26);

\path[draw=drawColor,line width= 0.4pt,line join=round,line cap=round] ( 49.20, 75.70) -- ( 43.20, 75.70);

\path[draw=drawColor,line width= 0.4pt,line join=round,line cap=round] ( 49.20,107.21) -- ( 43.20,107.21);

\path[draw=drawColor,line width= 0.4pt,line join=round,line cap=round] ( 49.20,138.72) -- ( 43.20,138.72);

\path[draw=drawColor,line width= 0.4pt,line join=round,line cap=round] ( 49.20,170.24) -- ( 43.20,170.24);

\path[draw=drawColor,line width= 0.4pt,line join=round,line cap=round] ( 49.20,201.75) -- ( 43.20,201.75);

\path[draw=drawColor,line width= 0.4pt,line join=round,line cap=round] ( 49.20,233.26) -- ( 43.20,233.26);

\node[text=drawColor,rotate= 90.00,anchor=base,inner sep=0pt, outer sep=0pt, scale=  1.00] at ( 34.80, 75.70) {0.0};

\node[text=drawColor,rotate= 90.00,anchor=base,inner sep=0pt, outer sep=0pt, scale=  1.00] at ( 34.80,107.21) {0.2};

\node[text=drawColor,rotate= 90.00,anchor=base,inner sep=0pt, outer sep=0pt, scale=  1.00] at ( 34.80,138.72) {0.4};

\node[text=drawColor,rotate= 90.00,anchor=base,inner sep=0pt, outer sep=0pt, scale=  1.00] at ( 34.80,170.24) {0.6};

\node[text=drawColor,rotate= 90.00,anchor=base,inner sep=0pt, outer sep=0pt, scale=  1.00] at ( 34.80,201.75) {0.8};

\node[text=drawColor,rotate= 90.00,anchor=base,inner sep=0pt, outer sep=0pt, scale=  1.00] at ( 34.80,233.26) {1.0};
\node[text=drawColor,rotate= 0.00,anchor=base,inner sep=0pt, outer sep=0pt, scale=  1.50] at ( 160.80,233.26) {$i=98$, $j=1$, $y=4$, $\omega=0.336$};
\end{scope}
\begin{scope}
\path[clip] ( 49.20, 61.20) rectangle (300.01,239.88);
\definecolor{drawColor}{RGB}{0,0,0}

\path[draw=drawColor,line width= 0.4pt,dash pattern=on 4pt off 4pt ,line join=round,line cap=round] ( 91.67, 75.70) -- ( 91.67, 82.99);

\path[draw=drawColor,line width= 0.4pt,dash pattern=on 4pt off 4pt ,line join=round,line cap=round] (124.84, 75.70) -- (124.84, 87.16);

\path[draw=drawColor,line width= 0.4pt,dash pattern=on 4pt off 4pt ,line join=round,line cap=round] (158.02, 75.70) -- (158.02,132.62);

\path[draw=drawColor,line width= 0.4pt,dash pattern=on 4pt off 4pt ,line join=round,line cap=round] (191.20, 75.70) -- (191.20,145.26);

\path[draw=drawColor,line width= 0.4pt,dash pattern=on 4pt off 4pt ,line join=round,line cap=round] (224.37, 75.70) -- (224.37, 80.49);

\path[draw=drawColor,line width= 0.4pt,dash pattern=on 4pt off 4pt ,line join=round,line cap=round] (257.55, 75.70) -- (257.55, 83.22);

\path[draw=drawColor,line width= 0.4pt,line join=round,line cap=round] (129.77, 75.70) --
	(130.65,109.25) --
	(131.53,116.02) --
	(132.41,120.47) --
	(133.29,123.87) --
	(134.17,126.67) --
	(135.05,129.08) --
	(135.93,131.20) --
	(136.81,133.11) --
	(137.69,134.86) --
	(138.57,136.49) --
	(139.45,138.01) --
	(140.33,139.45) --
	(141.21,140.81) --
	(142.10,142.12) --
	(142.98,143.38) --
	(143.86,144.60) --
	(144.74,145.78) --
	(145.62,146.93) --
	(146.50,148.06) --
	(147.38,149.17) --
	(148.26,150.26) --
	(149.14,151.34) --
	(150.02,152.41) --
	(150.90,153.48) --
	(151.78,154.54) --
	(152.66,155.60) --
	(153.54,156.67) --
	(154.42,157.74) --
	(155.30,158.82) --
	(156.18,159.92) --
	(157.06,161.03) --
	(157.94,162.16) --
	(158.82,163.32) --
	(159.70,164.50) --
	(160.58,165.72) --
	(161.47,166.99) --
	(162.35,168.30) --
	(163.23,169.68) --
	(164.11,171.12) --
	(164.99,172.66) --
	(165.87,174.30) --
	(166.75,176.06) --
	(167.63,178.00) --
	(168.51,180.15) --
	(169.39,182.59) --
	(170.27,185.45) --
	(171.15,188.95) --
	(172.03,193.58) --
	(172.91,200.83) --
	(173.79,215.01) --
	(174.67,198.73) --
	(175.55,192.38) --
	(176.43,188.10) --
	(177.31,184.79) --
	(178.19,182.04) --
	(179.07,179.68) --
	(179.95,177.59) --
	(180.84,175.70) --
	(181.72,173.97) --
	(182.60,172.36) --
	(183.48,170.85) --
	(184.36,169.43) --
	(185.24,168.07) --
	(186.12,166.77) --
	(187.00,165.52) --
	(187.88,164.31) --
	(188.76,163.13) --
	(189.64,161.99) --
	(190.52,160.86) --
	(191.40,159.76) --
	(192.28,158.67) --
	(193.16,157.60) --
	(194.04,156.53) --
	(194.92,155.47) --
	(195.80,154.42) --
	(196.68,153.36) --
	(197.56,152.30) --
	(198.44,151.23) --
	(199.32,150.16) --
	(200.21,149.07) --
	(201.09,147.96) --
	(201.97,146.84) --
	(202.85,145.69) --
	(203.73,144.51) --
	(204.61,143.29) --
	(205.49,142.04) --
	(206.37,140.73) --
	(207.25,139.37) --
	(208.13,137.93) --
	(209.01,136.41) --
	(209.89,134.79) --
	(210.77,133.04) --
	(211.65,131.13) --
	(212.53,129.01) --
	(213.41,126.61) --
	(214.29,123.82) --
	(215.17,120.42) --
	(216.05,115.97) --
	(216.93,109.21) --
	(217.81, 75.70);
\definecolor{drawColor}{RGB}{255,0,0}
\definecolor{fillColor}{RGB}{255,0,0}

\path[draw=drawColor,line width= 1.6pt,line join=round,line cap=round,fill=fillColor] (191.20, 75.70) circle (  1.50);
\end{scope}
\end{tikzpicture}

\begin{tikzpicture}[x=1pt,y=1pt]
\definecolor{fillColor}{RGB}{255,255,255}
\path[use as bounding box,fill=fillColor,fill opacity=0.00] (0,0) rectangle (325.21,289.08);
\begin{scope}
\path[clip] ( 49.20, 61.20) rectangle (300.01,239.88);
\definecolor{drawColor}{RGB}{0,0,0}

\path[draw=drawColor,line width= 0.4pt,line join=round,line cap=round] ( 91.67, 76.92) circle (  4.50);

\path[draw=drawColor,line width= 0.4pt,line join=round,line cap=round] (124.84, 77.39) circle (  4.50);

\path[draw=drawColor,line width= 0.4pt,line join=round,line cap=round] (158.02, 99.66) circle (  4.50);

\path[draw=drawColor,line width= 0.4pt,line join=round,line cap=round] (191.20,192.78) circle (  4.50);

\path[draw=drawColor,line width= 0.4pt,line join=round,line cap=round] (224.37, 81.41) circle (  4.50);

\path[draw=drawColor,line width= 0.4pt,line join=round,line cap=round] (257.55, 83.58) circle (  4.50);
\end{scope}
\begin{scope}
\path[clip] (  0.00,  0.00) rectangle (325.21,289.08);
\definecolor{drawColor}{RGB}{0,0,0}

\path[draw=drawColor,line width= 0.4pt,line join=round,line cap=round] ( 58.49, 61.20) -- (290.73, 61.20);

\path[draw=drawColor,line width= 0.4pt,line join=round,line cap=round] ( 58.49, 61.20) -- ( 58.49, 55.20);

\path[draw=drawColor,line width= 0.4pt,line join=round,line cap=round] ( 91.67, 61.20) -- ( 91.67, 55.20);

\path[draw=drawColor,line width= 0.4pt,line join=round,line cap=round] (124.84, 61.20) -- (124.84, 55.20);

\path[draw=drawColor,line width= 0.4pt,line join=round,line cap=round] (158.02, 61.20) -- (158.02, 55.20);

\path[draw=drawColor,line width= 0.4pt,line join=round,line cap=round] (191.20, 61.20) -- (191.20, 55.20);

\path[draw=drawColor,line width= 0.4pt,line join=round,line cap=round] (224.37, 61.20) -- (224.37, 55.20);

\path[draw=drawColor,line width= 0.4pt,line join=round,line cap=round] (257.55, 61.20) -- (257.55, 55.20);

\path[draw=drawColor,line width= 0.4pt,line join=round,line cap=round] (290.73, 61.20) -- (290.73, 55.20);

\node[text=drawColor,anchor=base,inner sep=0pt, outer sep=0pt, scale=  1.00] at ( 58.49, 39.60) {0};

\node[text=drawColor,anchor=base,inner sep=0pt, outer sep=0pt, scale=  1.00] at ( 91.67, 39.60) {1};

\node[text=drawColor,anchor=base,inner sep=0pt, outer sep=0pt, scale=  1.00] at (124.84, 39.60) {2};

\node[text=drawColor,anchor=base,inner sep=0pt, outer sep=0pt, scale=  1.00] at (158.02, 39.60) {3};

\node[text=drawColor,anchor=base,inner sep=0pt, outer sep=0pt, scale=  1.00] at (191.20, 39.60) {4};

\node[text=drawColor,anchor=base,inner sep=0pt, outer sep=0pt, scale=  1.00] at (224.37, 39.60) {5};

\node[text=drawColor,anchor=base,inner sep=0pt, outer sep=0pt, scale=  1.00] at (257.55, 39.60) {6};

\node[text=drawColor,anchor=base,inner sep=0pt, outer sep=0pt, scale=  1.00] at (290.73, 39.60) {7};

\path[draw=drawColor,line width= 0.4pt,line join=round,line cap=round] ( 49.20, 75.70) -- ( 49.20,233.26);

\path[draw=drawColor,line width= 0.4pt,line join=round,line cap=round] ( 49.20, 75.70) -- ( 43.20, 75.70);

\path[draw=drawColor,line width= 0.4pt,line join=round,line cap=round] ( 49.20,107.21) -- ( 43.20,107.21);

\path[draw=drawColor,line width= 0.4pt,line join=round,line cap=round] ( 49.20,138.72) -- ( 43.20,138.72);

\path[draw=drawColor,line width= 0.4pt,line join=round,line cap=round] ( 49.20,170.24) -- ( 43.20,170.24);

\path[draw=drawColor,line width= 0.4pt,line join=round,line cap=round] ( 49.20,201.75) -- ( 43.20,201.75);

\path[draw=drawColor,line width= 0.4pt,line join=round,line cap=round] ( 49.20,233.26) -- ( 43.20,233.26);

\node[text=drawColor,rotate= 90.00,anchor=base,inner sep=0pt, outer sep=0pt, scale=  1.00] at ( 34.80, 75.70) {0.0};

\node[text=drawColor,rotate= 90.00,anchor=base,inner sep=0pt, outer sep=0pt, scale=  1.00] at ( 34.80,107.21) {0.2};

\node[text=drawColor,rotate= 90.00,anchor=base,inner sep=0pt, outer sep=0pt, scale=  1.00] at ( 34.80,138.72) {0.4};

\node[text=drawColor,rotate= 90.00,anchor=base,inner sep=0pt, outer sep=0pt, scale=  1.00] at ( 34.80,170.24) {0.6};

\node[text=drawColor,rotate= 90.00,anchor=base,inner sep=0pt, outer sep=0pt, scale=  1.00] at ( 34.80,201.75) {0.8};

\node[text=drawColor,rotate= 90.00,anchor=base,inner sep=0pt, outer sep=0pt, scale=  1.00] at ( 34.80,233.26) {1.0};
\node[text=drawColor,rotate= 0.00,anchor=base,inner sep=0pt, outer sep=0pt, scale=  1.50] at ( 160.80,233.26) {$i=4$, $j=4$, $y=4$, $\omega=0.579$};
\end{scope}
\begin{scope}
\path[clip] ( 49.20, 61.20) rectangle (300.01,239.88);
\definecolor{drawColor}{RGB}{0,0,0}

\path[draw=drawColor,line width= 0.4pt,dash pattern=on 4pt off 4pt ,line join=round,line cap=round] ( 91.67, 75.70) -- ( 91.67, 76.92);

\path[draw=drawColor,line width= 0.4pt,dash pattern=on 4pt off 4pt ,line join=round,line cap=round] (124.84, 75.70) -- (124.84, 77.39);

\path[draw=drawColor,line width= 0.4pt,dash pattern=on 4pt off 4pt ,line join=round,line cap=round] (158.02, 75.70) -- (158.02, 99.66);

\path[draw=drawColor,line width= 0.4pt,dash pattern=on 4pt off 4pt ,line join=round,line cap=round] (191.20, 75.70) -- (191.20,192.78);

\path[draw=drawColor,line width= 0.4pt,dash pattern=on 4pt off 4pt ,line join=round,line cap=round] (224.37, 75.70) -- (224.37, 81.41);

\path[draw=drawColor,line width= 0.4pt,dash pattern=on 4pt off 4pt ,line join=round,line cap=round] (257.55, 75.70) -- (257.55, 83.58);

\path[draw=drawColor,line width= 0.4pt,line join=round,line cap=round] (165.48, 75.70) --
(166.05, 90.34) --
(166.62, 96.82) --
(167.19,101.75) --
(167.75,105.87) --
(168.32,109.47) --
(168.89,112.71) --
(169.45,115.68) --
(170.02,118.43) --
(170.59,121.02) --
(171.16,123.47) --
(171.72,125.80) --
(172.29,128.03) --
(172.86,130.19) --
(173.42,132.27) --
(173.99,134.30) --
(174.56,136.27) --
(175.12,138.19) --
(175.69,140.08) --
(176.26,141.94) --
(176.83,143.77) --
(177.39,145.58) --
(177.96,147.37) --
(178.53,149.14) --
(179.09,150.91) --
(179.66,152.67) --
(180.23,154.42) --
(180.80,156.17) --
(181.36,157.93) --
(181.93,159.69) --
(182.50,161.47) --
(183.06,163.26) --
(183.63,165.06) --
(184.20,166.89) --
(184.76,168.75) --
(185.33,170.63) --
(185.90,172.56) --
(186.47,174.53) --
(187.03,176.55) --
(187.60,178.63) --
(188.17,180.78) --
(188.73,183.01) --
(189.30,185.33) --
(189.87,187.77) --
(190.43,190.35) --
(191.00,193.09) --
(191.57,196.04) --
(192.14,199.26) --
(192.70,202.83) --
(193.27,206.91) --
(193.84,211.77) --
(194.40,218.10) --
(194.97,229.98) --
(195.54,218.52) --
(196.11,211.60) --
(196.67,206.39) --
(197.24,202.07) --
(197.81,198.29) --
(198.37,194.90) --
(198.94,191.80) --
(199.51,188.91) --
(200.07,186.21) --
(200.64,183.64) --
(201.21,181.20) --
(201.78,178.85) --
(202.34,176.59) --
(202.91,174.40) --
(203.48,172.26) --
(204.04,170.18) --
(204.61,168.14) --
(205.18,166.14) --
(205.74,164.16) --
(206.31,162.21) --
(206.88,160.28) --
(207.45,158.36) --
(208.01,156.45) --
(208.58,154.54) --
(209.15,152.64) --
(209.71,150.73) --
(210.28,148.81) --
(210.85,146.88) --
(211.42,144.93) --
(211.98,142.96) --
(212.55,140.96) --
(213.12,138.92) --
(213.68,136.84) --
(214.25,134.71) --
(214.82,132.52) --
(215.38,130.26) --
(215.95,127.92) --
(216.52,125.48) --
(217.09,122.93) --
(217.65,120.23) --
(218.22,117.36) --
(218.79,114.28) --
(219.35,110.91) --
(219.92,107.16) --
(220.49,102.88) --
(221.05, 97.75) --
(221.62, 91.00) --
(222.19, 75.70);
\definecolor{drawColor}{RGB}{255,0,0}
\definecolor{fillColor}{RGB}{255,0,0}

\path[draw=drawColor,line width= 1.6pt,line join=round,line cap=round,fill=fillColor] (191.20, 75.70) circle (  1.50);
\end{scope}
\end{tikzpicture}

%% file: fig4.tex
% Created by tikzDevice version 0.12.3.1 on 2021-01-10 19:30:05
% !TEX encoding = UTF-8 Unicode
\begin{tikzpicture}[x=1pt,y=1pt]
\definecolor{fillColor}{RGB}{255,255,255}
\path[use as bounding box,fill=fillColor,fill opacity=0.00] (0,0) rectangle (325.21,289.08);
\begin{scope}
\path[clip] ( 49.20, 61.20) rectangle (300.01,239.88);
\definecolor{drawColor}{RGB}{255,160,122}

\path[draw=drawColor,line width= 1.0pt,line join=round,line cap=round] (165.13, 67.82) --
	(165.79, 87.58) --
	(166.45, 94.76) --
	(167.11, 99.96) --
	(167.77,104.18) --
	(168.43,107.79) --
	(169.09,110.99) --
	(169.75,113.88) --
	(170.40,116.54) --
	(171.06,119.01) --
	(171.72,121.33) --
	(172.38,123.52) --
	(173.04,125.61) --
	(173.70,127.62) --
	(174.36,129.55) --
	(175.02,131.42) --
	(175.68,133.23) --
	(176.34,134.99) --
	(177.00,136.72) --
	(177.66,138.41) --
	(178.32,140.07) --
	(178.98,141.71) --
	(179.64,143.33) --
	(180.30,144.93) --
	(180.96,146.52) --
	(181.62,148.11) --
	(182.28,149.68) --
	(182.94,151.26) --
	(183.60,152.83) --
	(184.26,154.42) --
	(184.92,156.01) --
	(185.58,157.61) --
	(186.24,159.22) --
	(186.90,160.86) --
	(187.56,162.52) --
	(188.22,164.21) --
	(188.88,165.93) --
	(189.54,167.69) --
	(190.20,169.50) --
	(190.86,171.36) --
	(191.52,173.29) --
	(192.18,175.29) --
	(192.84,177.37) --
	(193.50,179.56) --
	(194.16,181.86) --
	(194.82,184.32) --
	(195.48,186.96) --
	(196.14,189.83) --
	(196.80,192.99) --
	(197.46,196.56) --
	(198.12,200.72) --
	(198.78,205.81) --
	(199.44,212.73) --
	(200.10,227.00) --
	(200.76,213.21) --
	(201.42,205.38) --
	(202.08,199.79) --
	(202.74,195.30) --
	(203.40,191.46) --
	(204.06,188.07) --
	(204.71,185.00) --
	(205.37,182.18) --
	(206.03,179.56) --
	(206.69,177.10) --
	(207.35,174.76) --
	(208.01,172.52) --
	(208.67,170.38) --
	(209.33,168.31) --
	(209.99,166.30) --
	(210.65,164.35) --
	(211.31,162.44) --
	(211.97,160.56) --
	(212.63,158.71) --
	(213.29,156.89) --
	(213.95,155.09) --
	(214.61,153.29) --
	(215.27,151.51) --
	(215.93,149.73) --
	(216.59,147.94) --
	(217.25,146.15) --
	(217.91,144.35) --
	(218.57,142.53) --
	(219.23,140.68) --
	(219.89,138.81) --
	(220.55,136.90) --
	(221.21,134.95) --
	(221.87,132.95) --
	(222.53,130.88) --
	(223.19,128.75) --
	(223.85,126.52) --
	(224.51,124.19) --
	(225.17,121.74) --
	(225.83,119.13) --
	(226.49,116.34) --
	(227.15,113.29) --
	(227.81,109.94) --
	(228.47,106.14) --
	(229.13,101.72) --
	(229.79, 96.26) --
	(230.45, 88.72) --
	(231.11, 67.82);
\end{scope}
\begin{scope}
\path[clip] (  0.00,  0.00) rectangle (325.21,289.08);
\definecolor{drawColor}{RGB}{0,0,0}

\path[draw=drawColor,line width= 0.4pt,line join=round,line cap=round] ( 58.49, 61.20) -- (290.73, 61.20);

\path[draw=drawColor,line width= 0.4pt,line join=round,line cap=round] ( 58.49, 61.20) -- ( 58.49, 55.20);

\path[draw=drawColor,line width= 0.4pt,line join=round,line cap=round] ( 91.67, 61.20) -- ( 91.67, 55.20);

\path[draw=drawColor,line width= 0.4pt,line join=round,line cap=round] (124.84, 61.20) -- (124.84, 55.20);

\path[draw=drawColor,line width= 0.4pt,line join=round,line cap=round] (158.02, 61.20) -- (158.02, 55.20);

\path[draw=drawColor,line width= 0.4pt,line join=round,line cap=round] (191.20, 61.20) -- (191.20, 55.20);

\path[draw=drawColor,line width= 0.4pt,line join=round,line cap=round] (224.37, 61.20) -- (224.37, 55.20);

\path[draw=drawColor,line width= 0.4pt,line join=round,line cap=round] (257.55, 61.20) -- (257.55, 55.20);

\path[draw=drawColor,line width= 0.4pt,line join=round,line cap=round] (290.73, 61.20) -- (290.73, 55.20);

\node[text=drawColor,anchor=base,inner sep=0pt, outer sep=0pt, scale=  1.00] at ( 58.49, 39.60) {0};

\node[text=drawColor,anchor=base,inner sep=0pt, outer sep=0pt, scale=  1.00] at ( 91.67, 39.60) {1};

\node[text=drawColor,anchor=base,inner sep=0pt, outer sep=0pt, scale=  1.00] at (124.84, 39.60) {2};

\node[text=drawColor,anchor=base,inner sep=0pt, outer sep=0pt, scale=  1.00] at (158.02, 39.60) {3};

\node[text=drawColor,anchor=base,inner sep=0pt, outer sep=0pt, scale=  1.00] at (191.20, 39.60) {4};

\node[text=drawColor,anchor=base,inner sep=0pt, outer sep=0pt, scale=  1.00] at (224.37, 39.60) {5};

\node[text=drawColor,anchor=base,inner sep=0pt, outer sep=0pt, scale=  1.00] at (257.55, 39.60) {6};

\node[text=drawColor,anchor=base,inner sep=0pt, outer sep=0pt, scale=  1.00] at (290.73, 39.60) {7};

\path[draw=drawColor,line width= 0.4pt,line join=round,line cap=round] ( 49.20, 67.82) -- ( 49.20,233.26);

\path[draw=drawColor,line width= 0.4pt,line join=round,line cap=round] ( 49.20, 67.82) -- ( 43.20, 67.82);

\path[draw=drawColor,line width= 0.4pt,line join=round,line cap=round] ( 49.20,100.91) -- ( 43.20,100.91);

\path[draw=drawColor,line width= 0.4pt,line join=round,line cap=round] ( 49.20,134.00) -- ( 43.20,134.00);

\path[draw=drawColor,line width= 0.4pt,line join=round,line cap=round] ( 49.20,167.08) -- ( 43.20,167.08);

\path[draw=drawColor,line width= 0.4pt,line join=round,line cap=round] ( 49.20,200.17) -- ( 43.20,200.17);

\path[draw=drawColor,line width= 0.4pt,line join=round,line cap=round] ( 49.20,233.26) -- ( 43.20,233.26);

\node[text=drawColor,rotate= 90.00,anchor=base,inner sep=0pt, outer sep=0pt, scale=  1.00] at ( 34.80, 67.82) {0.0};

\node[text=drawColor,rotate= 90.00,anchor=base,inner sep=0pt, outer sep=0pt, scale=  1.00] at ( 34.80,100.91) {0.2};

\node[text=drawColor,rotate= 90.00,anchor=base,inner sep=0pt, outer sep=0pt, scale=  1.00] at ( 34.80,134.00) {0.4};

\node[text=drawColor,rotate= 90.00,anchor=base,inner sep=0pt, outer sep=0pt, scale=  1.00] at ( 34.80,167.08) {0.6};

\node[text=drawColor,rotate= 90.00,anchor=base,inner sep=0pt, outer sep=0pt, scale=  1.00] at ( 34.80,200.17) {0.8};

\node[text=drawColor,rotate= 90.00,anchor=base,inner sep=0pt, outer sep=0pt, scale=  1.00] at ( 34.80,233.26) {1.0};
\end{scope}
\begin{scope}
\path[clip] ( 49.20, 61.20) rectangle (300.01,239.88);
\definecolor{drawColor}{RGB}{96,123,139}

\path[draw=drawColor,line width= 1.0pt,line join=round,line cap=round] (132.74, 67.82) --
	(133.67,106.19) --
	(134.61,113.00) --
	(135.54,117.42) --
	(136.48,120.76) --
	(137.42,123.50) --
	(138.35,125.84) --
	(139.29,127.89) --
	(140.23,129.74) --
	(141.16,131.43) --
	(142.10,132.99) --
	(143.04,134.45) --
	(143.97,135.82) --
	(144.91,137.13) --
	(145.85,138.38) --
	(146.78,139.58) --
	(147.72,140.74) --
	(148.65,141.86) --
	(149.59,142.95) --
	(150.53,144.02) --
	(151.46,145.07) --
	(152.40,146.11) --
	(153.34,147.13) --
	(154.27,148.14) --
	(155.21,149.14) --
	(156.15,150.14) --
	(157.08,151.14) --
	(158.02,152.14) --
	(158.96,153.14) --
	(159.89,154.15) --
	(160.83,155.17) --
	(161.76,156.21) --
	(162.70,157.26) --
	(163.64,158.34) --
	(164.57,159.44) --
	(165.51,160.57) --
	(166.45,161.73) --
	(167.38,162.94) --
	(168.32,164.20) --
	(169.26,165.52) --
	(170.19,166.91) --
	(171.13,168.39) --
	(172.06,169.97) --
	(173.00,171.69) --
	(173.94,173.57) --
	(174.87,175.68) --
	(175.81,178.08) --
	(176.75,180.91) --
	(177.68,184.42) --
	(178.62,189.14) --
	(179.56,196.85) --
	(180.49,207.56) --
	(181.43,192.91) --
	(182.37,186.75) --
	(183.30,182.58) --
	(184.24,179.35) --
	(185.17,176.69) --
	(186.11,174.40) --
	(187.05,172.37) --
	(187.98,170.54) --
	(188.92,168.86) --
	(189.86,167.31) --
	(190.79,165.85) --
	(191.73,164.47) --
	(192.67,163.16) --
	(193.60,161.90) --
	(194.54,160.70) --
	(195.48,159.53) --
	(196.41,158.39) --
	(197.35,157.28) --
	(198.28,156.19) --
	(199.22,155.12) --
	(200.16,154.07) --
	(201.09,153.02) --
	(202.03,151.99) --
	(202.97,150.95) --
	(203.90,149.92) --
	(204.84,148.89) --
	(205.78,147.85) --
	(206.71,146.81) --
	(207.65,145.75) --
	(208.59,144.68) --
	(209.52,143.59) --
	(210.46,142.47) --
	(211.39,141.33) --
	(212.33,140.15) --
	(213.27,138.93) --
	(214.20,137.67) --
	(215.14,136.34) --
	(216.08,134.95) --
	(217.01,133.48) --
	(217.95,131.90) --
	(218.89,130.19) --
	(219.82,128.33) --
	(220.76,126.26) --
	(221.69,123.90) --
	(222.63,121.15) --
	(223.57,117.78) --
	(224.50,113.34) --
	(225.44,106.48) --
	(226.38, 67.82);
\definecolor{drawColor}{RGB}{180,82,205}

\path[draw=drawColor,line width= 1.0pt,line join=round,line cap=round] (156.04, 67.82) --
	(156.86, 99.99) --
	(157.68,107.12) --
	(158.50,111.87) --
	(159.32,115.53) --
	(160.14,118.54) --
	(160.96,121.14) --
	(161.78,123.44) --
	(162.60,125.51) --
	(163.42,127.41) --
	(164.24,129.17) --
	(165.06,130.82) --
	(165.88,132.37) --
	(166.70,133.85) --
	(167.52,135.26) --
	(168.34,136.62) --
	(169.16,137.93) --
	(169.98,139.20) --
	(170.80,140.44) --
	(171.62,141.65) --
	(172.44,142.83) --
	(173.26,143.99) --
	(174.08,145.14) --
	(174.90,146.27) --
	(175.72,147.39) --
	(176.54,148.50) --
	(177.36,149.61) --
	(178.18,150.71) --
	(179.00,151.82) --
	(179.82,152.93) --
	(180.64,154.04) --
	(181.46,155.16) --
	(182.27,156.30) --
	(183.09,157.45) --
	(183.91,158.61) --
	(184.73,159.80) --
	(185.55,161.02) --
	(186.37,162.27) --
	(187.19,163.55) --
	(188.01,164.87) --
	(188.83,166.25) --
	(189.65,167.68) --
	(190.47,169.18) --
	(191.29,170.76) --
	(192.11,172.44) --
	(192.93,174.24) --
	(193.75,176.19) --
	(194.57,178.33) --
	(195.39,180.71) --
	(196.21,183.42) --
	(197.03,186.60) --
	(197.85,190.53) --
	(198.67,195.80) --
	(199.49,204.45) --
	(200.31,209.66) --
	(201.13,196.93) --
	(201.95,190.59) --
	(202.77,186.10) --
	(203.59,182.55) --
	(204.41,179.56) --
	(205.23,176.97) --
	(206.05,174.65) --
	(206.87,172.54) --
	(207.69,170.60) --
	(208.50,168.78) --
	(209.32,167.07) --
	(210.14,165.44) --
	(210.96,163.89) --
	(211.78,162.39) --
	(212.60,160.95) --
	(213.42,159.54) --
	(214.24,158.17) --
	(215.06,156.83) --
	(215.88,155.51) --
	(216.70,154.20) --
	(217.52,152.91) --
	(218.34,151.62) --
	(219.16,150.34) --
	(219.98,149.06) --
	(220.80,147.77) --
	(221.62,146.48) --
	(222.44,145.17) --
	(223.26,143.84) --
	(224.08,142.49) --
	(224.90,141.11) --
	(225.72,139.69) --
	(226.54,138.23) --
	(227.36,136.72) --
	(228.18,135.14) --
	(229.00,133.49) --
	(229.82,131.75) --
	(230.64,129.90) --
	(231.46,127.90) --
	(232.28,125.74) --
	(233.10,123.34) --
	(233.92,120.64) --
	(234.73,117.51) --
	(235.55,113.72) --
	(236.37,108.80) --
	(237.19,101.41) --
	(238.01, 67.82);
\end{scope}
\begin{scope}
\path[clip] (  0.00,  0.00) rectangle (325.21,289.08);
\definecolor{fillColor}{RGB}{255,160,122}

\path[fill=fillColor] ( 67.86, 15.00) rectangle ( 75.06,  9.00);
\definecolor{fillColor}{RGB}{96,123,139}

\path[fill=fillColor] (118.61, 15.00) rectangle (125.81,  9.00);
\definecolor{fillColor}{RGB}{180,82,205}

\path[fill=fillColor] (169.36, 15.00) rectangle (176.56,  9.00);
\definecolor{drawColor}{RGB}{0,0,0}

\node[text=drawColor,anchor=base west,inner sep=0pt, outer sep=0pt, scale=  1.30] at ( 84.06,  8.56) {i=4};

\node[text=drawColor,anchor=base west,inner sep=0pt, outer sep=0pt, scale=  1.30] at (134.81,  8.56) {i=28};

\node[text=drawColor,anchor=base west,inner sep=0pt, outer sep=0pt, scale=  1.30] at (185.56,  8.56) {i=321};
\end{scope}
\end{tikzpicture}